\documentclass[aps, prl, twocolumn,nofootinbib,superscriptaddress]{revtex4-2}

\usepackage[mathcal]{euscript}
\usepackage{amsmath}
\usepackage{textcomp, gensymb}
\usepackage[version=3]{mhchem} % Formula subscripts using \ce{}
\usepackage{amssymb}
\usepackage[dvipsnames]{xcolor}
\usepackage{graphicx}
\usepackage[normalem]{ulem} % for strikout text \sout{}
\usepackage{xr}
\usepackage[left]{lineno} 
\usepackage{hyperref}
\usepackage{dsfont}

\usepackage{physics} % for bra ket notation in the SI (ko4)

\usepackage[version=3]{mhchem} % Formula subscripts using \ce{}
\makeatletter
\DeclareUnicodeCharacter{2212}{\textendash}
\newcommand*{\addFileDependency}[1]{% argument=file name and extension
\typeout{(#1)}% latexmk will find this if $recorder=0
\@addtofilelist{#1}
\newcommand{\beq}{\begin{equation}}
	\newcommand{\eeq}{\end{equation}}

\IfFileExists{#1}{}{\typeout{No file #1.}}
}\makeatother

\begin{document}
%\linenumbers
\title{Non-Hermitian topology in the quantum Hall effect of graphene}

% \author{Burak \"Ozer}[1,2]
% \author{Kyrylo Ochkan}[1,2]
% \author{Romain Giraud}[1,3]

% \affiliation[1]{Leibniz Institute for Solid State and Materials Research,
% IFW Dresden, Helmholtzstrasse 20, 01069 Dresden, Germany}
% \affiliation[2]{W\"{u}rzburg-Dresden Cluster of Excellence ct.qmat, 01062 Dresden, Germany}
% \affiliation[3]{Université Grenoble Alpes, CNRS, CEA, Grenoble-INP, Spintec, F-38000 Grenoble, France}

\author{Burak \"Ozer$^*$}
\affiliation{Leibniz Institute for Solid State and Materials Research,
IFW Dresden, Helmholtzstrasse 20, 01069 Dresden, Germany}
\affiliation{W\"{u}rzburg-Dresden Cluster of Excellence ct.qmat, 01062 Dresden, Germany}

\author{Kyrylo Ochkan$^*$}
\affiliation{Leibniz Institute for Solid State and Materials Research,
IFW Dresden, Helmholtzstrasse 20, 01069 Dresden, Germany}
\affiliation{W\"{u}rzburg-Dresden Cluster of Excellence ct.qmat, 01062 Dresden, Germany}

\author{Raghav Chaturvedi}
\affiliation{Leibniz Institute for Solid State and Materials Research,
IFW Dresden, Helmholtzstrasse 20, 01069 Dresden, Germany}
\affiliation{W\"{u}rzburg-Dresden Cluster of Excellence ct.qmat, 01062 Dresden, Germany}

\author{Evgenii Maltsev}
\affiliation{Leibniz Institute for Solid State and Materials Research,
IFW Dresden, Helmholtzstrasse 20, 01069 Dresden, Germany}
\affiliation{W\"{u}rzburg-Dresden Cluster of Excellence ct.qmat, 01062 Dresden, Germany}

\author{Viktor K\"{o}nye}
\affiliation{Institute for Theoretical Physics Amsterdam,  University of Amsterdam, Science Park 904, 1098 XH Amsterdam, The Netherlands}

\author{Romain Giraud}
\affiliation{Leibniz Institute for Solid State and Materials Research,
IFW Dresden, Helmholtzstrasse 20, 01069 Dresden, Germany}
\affiliation{Université Grenoble Alpes, CNRS, CEA, Grenoble-INP, Spintec, F-38000 Grenoble, France}

\author{Arthur Veyrat}
\affiliation{Leibniz Institute for Solid State and Materials Research,
IFW Dresden, Helmholtzstrasse 20, 01069 Dresden, Germany}
\affiliation{W\"{u}rzburg-Dresden Cluster of Excellence ct.qmat, 01062 Dresden, Germany}

\author{Ewelina M. Hankiewicz}
\affiliation{W\"{u}rzburg-Dresden Cluster of Excellence ct.qmat, 01062 Dresden, Germany}
\affiliation{Institute for Theoretical Physics and Astrophysics, Julius-Maximilians-Universit\"{a}t W\"{u}rzburg, D-97074 W\"{u}rzburg, Germany}

\author{Kenji Watanabe}
\affiliation{Research Center for Electronic and Optical Materials, National Institute for Materials Science, 1-1 Namiki, Tsukuba 305-0044, Japan}

\author{Takashi Taniguchi}
\affiliation{Research Center for Materials Nanoarchitectonics, National Institute for Materials Science,  1-1 Namiki, Tsukuba 305-0044, Japan}

\author{Bernd B\"{u}chner}
\affiliation{Leibniz Institute for Solid State and Materials Research,
IFW Dresden, Helmholtzstrasse 20, 01069 Dresden, Germany}
\affiliation{W\"{u}rzburg-Dresden Cluster of Excellence ct.qmat, 01062 Dresden, Germany}
\affiliation{Department of Physics, TU Dresden, D-01062 Dresden, Germany}

\author{Jeroen van den Brink}
\affiliation{Leibniz Institute for Solid State and Materials Research,
IFW Dresden, Helmholtzstrasse 20, 01069 Dresden, Germany}
\affiliation{W\"{u}rzburg-Dresden Cluster of Excellence ct.qmat, 01062 Dresden, Germany}
\affiliation{Department of Physics, TU Dresden, D-01062 Dresden, Germany}

\author{Ion Cosma Fulga}
\affiliation{Leibniz Institute for Solid State and Materials Research,
IFW Dresden, Helmholtzstrasse 20, 01069 Dresden, Germany}
\affiliation{W\"{u}rzburg-Dresden Cluster of Excellence ct.qmat, 01062 Dresden, Germany}

\author{Joseph Dufouleur\thanks{j.dufouleur@ifw-dresden.de}}
\affiliation{Leibniz Institute for Solid State and Materials Research,
IFW Dresden, Helmholtzstrasse 20, 01069 Dresden, Germany}
\affiliation{W\"{u}rzburg-Dresden Cluster of Excellence ct.qmat, 01062 Dresden, Germany}

\author{Louis Veyrat}\thanks{louis.veyrat@lncmi.cnrs.fr}
\affiliation{Leibniz Institute for Solid State and Materials Research,
IFW Dresden, Helmholtzstrasse 20, 01069 Dresden, Germany}
\affiliation{W\"{u}rzburg-Dresden Cluster of Excellence ct.qmat, 01062 Dresden, Germany}
\affiliation{CNRS, Laboratoire National des Champs Magnétiques Intenses, Université Grenoble-Alpes, Université Toulouse 3, INSA-Toulouse, EMFL, 31400 Toulouse, France}

\date{\today}

%\twocolumn
\begin{abstract}
Quantum Hall phases have recently emerged as a platform to investigate non-Hermitian topology in condensed-matter systems. 
This platform is particularly interesting due to its tunability, which allows to modify the properties and topology of the investigated non-Hermitian phases by tuning external parameters of the system such as the magnetic field. 
Here, we show the tunability of non-Hermitian topology’s chirality in a graphene heterostructure using a gate voltage. By changing the charge carrier density, we unveil some novel properties specific to different quantum Hall regimes.
First, we find that the best quantization of the non-Hermitian topological invariant is interestingly obtained at very high filling factor rather than on well-quantized quantum Hall plateaus.
This is of particular importance for the efficient operation of devices based on non-Hermitian topology. 
Moreover, we observe an additional non-Hermitian topological phase in the insulating $\nu=0$ quantum Hall plateau, which survives at lower fields than the opening of the $\nu=0$ gap, confirming a recent prediction of a disorder-induced trivial phase.
%This observation confirms a recent prediction of an additional non-Hermitian topological phase that can be observed in the conductance matrix of graphene, caused by imperfections and disorder.
Our results evidence graphene as a promising platform for the study of non-Hermitian physics and of emergent phases in such topological devices.% in particular non-Hermitian topology.
\end{abstract}
\maketitle
\def\thefootnote{*}\footnotetext{These authors contributed equally to this work}\def\thefootnote{\arabic{footnote}}

%\section{Introduction}
Non-Hermitian quantum physics describes open systems where energy or particles can be added or removed, and is associated with gains and losses \cite{FoaTorres2019, Ashida2020, Bergholtz2021}. 
Similar to Hermitian systems, non-Hermitian systems can also have topological properties, so-called non-Hermitian topology~\cite{Bergholtz2021}. 
These non-Hermitian topological phases possess robust properties which can be used in concrete applications, such as exponentially precise sensors \cite{Budich2020, Konye2024, Parto2023, Yuan2023}, amplifiers \cite{Wang2022}, and light funnels \cite{Weidemann2020}.
In condensed matter, one way to induce non-Hermitian topology is to tailor interactions to induce gains and losses, which remains very challenging. An alternative consists of considering a chiral Hermitian system coupled to a dissipative environment \cite{Lee2019}.
Although closed systems, such as the quantum Hall (QH) effect, are generally described as Hermitian phases, coupling %Although Hermitian descriptions are generally used for closed systems, such as the quantum Hall effect, coupling 
those systems to an environment induces gain and losses which are correctly captured by an effective non-Hermitian description~\cite{Nakamura2023, Schindler2023,Lee2019}.
%In condensed matter systems, although gains and losses can be introduced through coupling with the environment, their fine-tuning to produce non-trivial phases remains a challenge.

\begin{figure*}[t]
\centering
\includegraphics[width=1.\textwidth]{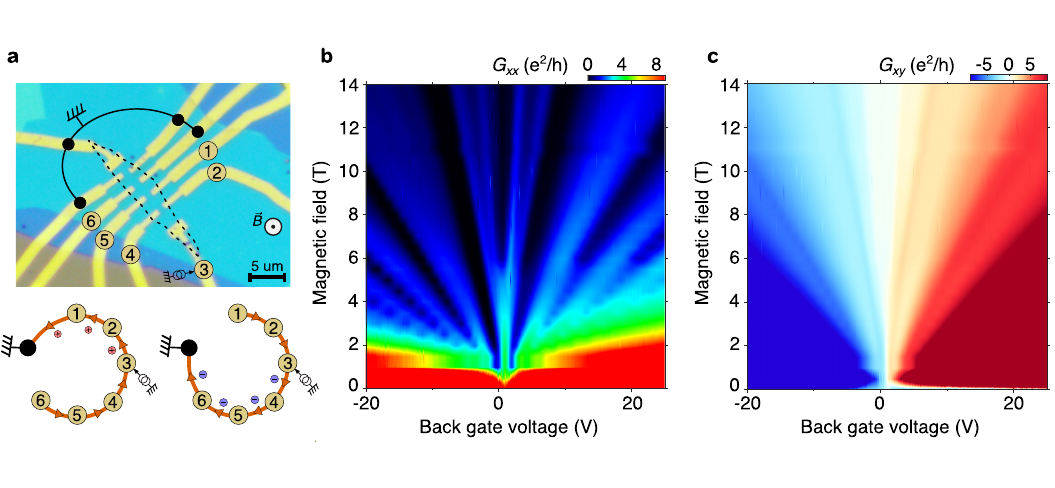}
\caption{
\textbf{Device characterization. a}: 
Optical microscopy picture of the contacted hBN/Graphene/hBN heterostructure with numbered contacts. 
The graphene shape is highlighted by a black dashed line. 
All contacts after contact 6 are grounded as depicted by the black line. 
The effective corresponding Hatano-Nelson chain is shown schematically below the picture, for both chiralities of the quantum Hall edge channels (electron and hole charge carriers depicted by - and + symbols), for the case where the current source is on contact 3.
\textbf{b, c}: fan diagrams of the longitudinal conductance $G_{xx}$ and of the transverse conductance $G_{xy}$ of the device, respectively, with respect to magnetic field and back-gate voltage.
}
\label{Fig:characterization}
\end{figure*}

Recently, a condensed-matter platform for the realisation and study of non-Hermitian topology was demonstrated in a QH device \cite{Ochkan2024}. 
Although the QH phase is intrinsically Hermitian, the chiral propagation of edge channels in a multi-terminal QH device closely resembles the description of a one-dimensional (1D) non-Hermitian chain with directional hopping \cite{Franca2021} [see Fig.~\ref{Fig:characterization}(a)], known as the Hatano-Nelson (HN) model~\cite{Hatano1996}. 
The HN Hamiltonian,
\begin{equation}\label{eq:HNHam}
{\cal H}_{\rm HN} = \sum_j J_{\rm left} c^{\dag}_{j-1} c_j + J_{\rm right} c^{\dag}_{j+1} c_j = \mathbf{c}^{\dag} H_{\rm HN} \mathbf{c}
\end{equation}
describes a 1D chain on which quantum particles hop between neighboring sites (site index $j$, creation operator $c^\dag_j$), where the hopping to the left and to the right are real numbers with different magnitudes, $J_{\rm left} \neq J_{\rm right}$. 
$\mathbf{c}$ is a column vector formed from all the annihilation operators and $H_{\rm HN}$ is the Hamiltonian matrix, whose size is given by the number of sites in the chain. 
If $J_{\rm left} \neq J_{\rm right}$, a net current of particles will be induced in a given direction, resulting in a given chirality (left/right) and in non-Hermiticity~\cite{Zhang2020}.
The Hamiltonian $H_{\rm HN}$ can be linked to the QH effect through a QH device’s conductance matrix. 
Indeed, the conductance matrix of a perfect QH system with right chirality maps onto the HN Hamiltonian matrix $H_{\rm HN}$ with $J_{\rm left} =0$ and $J_{\rm right} = -1$, making it possible to study the topology of the non-Hermitian HN model through the conductance matrix of a multi-connected QH device~\cite{Ochkan2024}.
%In a perfect QH system with right-chirality, $J_{\rm left} =0$ and $J_{\rm right} =1$, and the conductance matrix $G_{ij}$ directly maps the HN Hamitlonian, making it possible to study the topology of the non-Hermitian HN model through the conductance matrix of a simple quantum Hall device\cite{Ochkan2024}.
    
The QH platform is particularly interesting to study non-Hermitian topology thanks to its tunability by external parameters. 
Key parameters of the HN model (transmission coefficients, presence of non-diagonal terms linking non-adjacent sites, open/periodic boundary conditions etc) are related to properties of the QH system (number of edge channels, position of the chemical potential on a plateau or in-between plateaus, geometry of the QH device etc) that can be tuned through external parameters such as the magnetic field or an electrostatic gate.
In AlGaAs two-dimensional electron gases (2DEG), on which the first report of non-Hermitian topology studied in a QH system~\cite{Ochkan2024} was based, the carrier type cannot, however, be continuously tuned from electrons to holes to reverse the chirality and thus switch the topological state. 
Switching topology then requires to reverse the magnetic field, thus first closing the cyclotron gap at zero field.
%In AlGaAs two-dimensional electron gases (2DEG), on which the first report of non-Hermitian topology studied in a QH system~\cite{Ochkan2024} was based, changing the non-Hermitian topology (left/right chirality) however requires to reverse the magnetic field and therefore to transition through a topologically trivial phase at zero magnetic field.
    
In this work, we experimentally explore the non-Hermitian topology of the conductance matrix in the QH regime of monolayer graphene.
%the topology of the HN model via the study of the QH regime of monolayer graphene. 
Contrary to AlGaAs-based 2DEGs, graphene can easily be tuned between electron-like and hole-like doping at stable temperature and magnetic field by applying a gate voltage on a single device, offering some additional tunability of the non-Hermitian topology. 
Similar to what was observed recently in AlGaAs, we confirm that the conductance matrix of a graphene-based QH device maps to the Hamiltonian of the HN model. 
Between the hole and the electron sides, we observe as expected a transition between the left-chiral and right-chiral non-Hermitian topology, associated with a non-Hermitian topological invariant $w_{\rm PD} = \pm 1$ (see Supplementary Information. 
We moreover observe an additional non-Hermitian phase, associated with an invariant $w_{\rm PD} = 0$, in the insulating, $\nu = 0$ QH phase of graphene. 
This non-Hermitian phase survives even beyond the $\nu = 0$ phase, which according to a recent theoretical prediction\cite{Chaturvedi2024} is intrinsic to the existence of disorder in the device.
\\ \\
\begin{figure*}[t]
\centering
\includegraphics[width=1\textwidth]{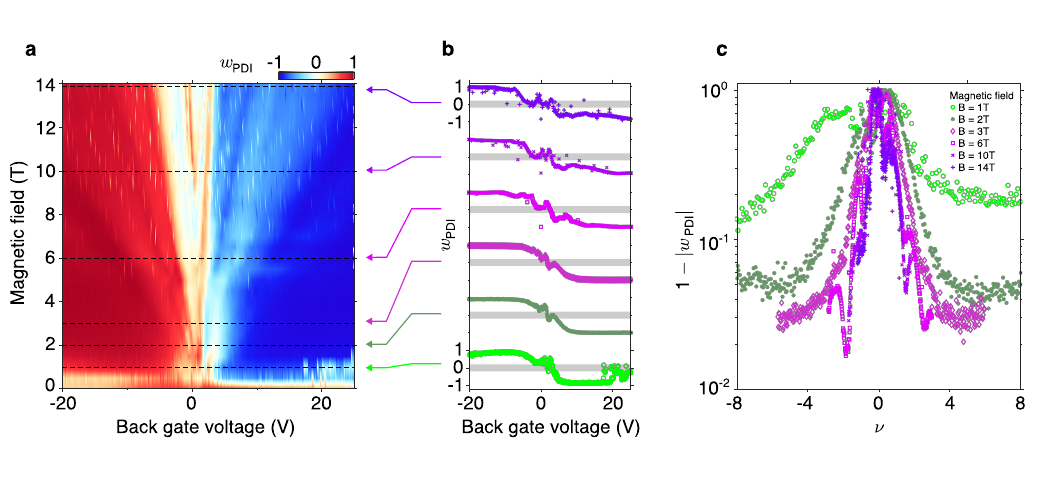}
\caption{
\textbf{Non-Hermitian topology in graphene. a}: %(\textcolor{blue}{(Raghav: Plot (a) shows the invariant going from -1 to 1 when we increase back gate. However the line cuts in (b) show it going from 1 to -1. Maybe an issue with the colorbar? Maybe a similar thing in S2 and S5 as well?)}
fan diagram of the non-Hermitian topological invariant $w_{\rm PD}$ calculated from the device's conductance matrices measured at each point of the fan diagram.
\textbf{b}: Line cuts of the invariant $w_{\rm PD}$ with respect to gate voltage $V_g$ at different magnetic fields $B$. 
The line cuts are shifted vertically for clarity. 
The grey areas correspond to the range $\pm0.2$ around zero. 
\textbf{c}: Deviation of $w_{\rm PD}$ from perfect quantization, measured as $1-|w_{\rm PD}|$, with respect to filling factor $\nu$ at different magnetic fields. 
The field values in panels \textbf{b} and \textbf{c} correspond to the dashed lines in panel \textbf{a}.
}
\label{Fig:wPDI}
\end{figure*}
%\section{Results}
    
The device studied is a monolayer graphene encapsulated in-between hBN flakes and deposited on a \ce{Si++}/\ce{SiO2} substrate using the dry van-der-Waals pickup technique~\cite{Zimmermann2017}. 
Contacts were taken by patterning trenches using e-beam lithography and then etching the heterostructure using a \ce{CHF3}/\ce{O2} plasma, before depositing Ti/Au contacts in the etched trenches~\cite{Veyrat2020}. 
The contacted device is pictured in Fig.~\ref{Fig:characterization}(a). 
A back-gate voltage $V_g$ applied to the conducting \ce{Si} substrate allows to tune the carrier density of graphene.

Upon applying a magnetic field $B$, graphene rapidly enters the QH regime. 
Figures \ref{Fig:characterization}(b) and (c) show the fan diagrams of the longitudinal and transverse conductances, respectively. 
The device enters the QH regime above $\sim 1$T, and degeneracy lifting of the $N=0$ Landau level occurs at $\sim 6$T and $\sim 8$T with the opening of the $\nu=0$ and $\nu=\pm 1$ plateaus, respectively.

The analogy between the HN model and the QH system occurs through the conductance matrix. 
Indeed, the conductance matrix of a QH device is directly related to the HN model Hamiltonian\cite{Ochkan2024}. 
In the QH regime, the contacts are connected to each other through 1D edge modes, forming a closed chain. 
By adding a contact drain, whose potential is fixed, the chain is effectively cut [Fig.~\ref{Fig:characterization}(a)], so that the circuit realizes the open boundary condition case of the HN model.

To study the topology of the HN model using our graphene device, we investigated the conductance matrix. 
To extract one column of the conductance matrix, the potential of each floating contact must be measured for a given position of the contact source (see Supplementary Information and Fig.~\ref{Fig:S_matrix_measurement}). 
For technical reasons we restricted the device geometry to a chain made of six terminals, the remaining contacts being all grounded, as pictured in Fig.~\ref{Fig:characterization}(a). 
In order to characterize the conductance matrix with respect to both the magnetic field and the chemical potential, we measured the fan diagram of the potential of each of the six contacts, once for each of the six possible positions of the current source (using voltage polarization of the source contact at $400\mu V$). 
Combining all these 36 fan diagrams, we obtain the total resistance matrix $R$ and, by inverting $R$, the conductance matrix $G$ of the device (restricted to six independent contacts) for each point of the ($B$, $V_g$) map. 
As expected and previously reported for AlGaAs \cite{Ochkan2024}, the conductance matrices are found to be analogous to the Hamiltonian matrix of the Hatano-Nelson model $H_{HN}$ (see Supplementary Information and  Fig.~\ref{Fig:S_matrices}). 
This confirms graphene-based QH devices as a good platform to study non-Hermitian topology.

To investigate quantitatively non-Hermitian topology in our graphene device, we look at a topological invariant and at the non-Hermitian skin effect \cite{MartinezAlvarez2018, Yao2018, Lee2019b, Borgnia2020, Kawabata2019}. 
Using the same calculations as in Ref.~\cite{Ochkan2024} (see Supplementary Information), we use the measured conductance matrices to calculate, for each point of the ($B$, $V_g$) map, the winding number $w_{\rm PD}$ of the HN Hamiltonian corresponding to the conductance matrix at this point. 
$w_{\rm PD}$, a topological invariant based on the polar decomposition of the HN model, captures the topology (chiral left or chiral right) of the open finite HN chain~\cite{Hughes2021, Ochkan2024}. 
While the sign is associated with the chirality, the value is real and approaches quantized values of $\pm1$ deep in the topological phases, and can be calculated from the HN Hamiltonian (i.e. in our case from the conductance matrix). 
We obtain a fan diagram of the $w_{\rm PD}$ topological invariant of the QH graphene system, which is presented in Fig.~\ref{Fig:wPDI}(a).

\begin{figure*}[t]
\centering
\includegraphics[width=1\textwidth]{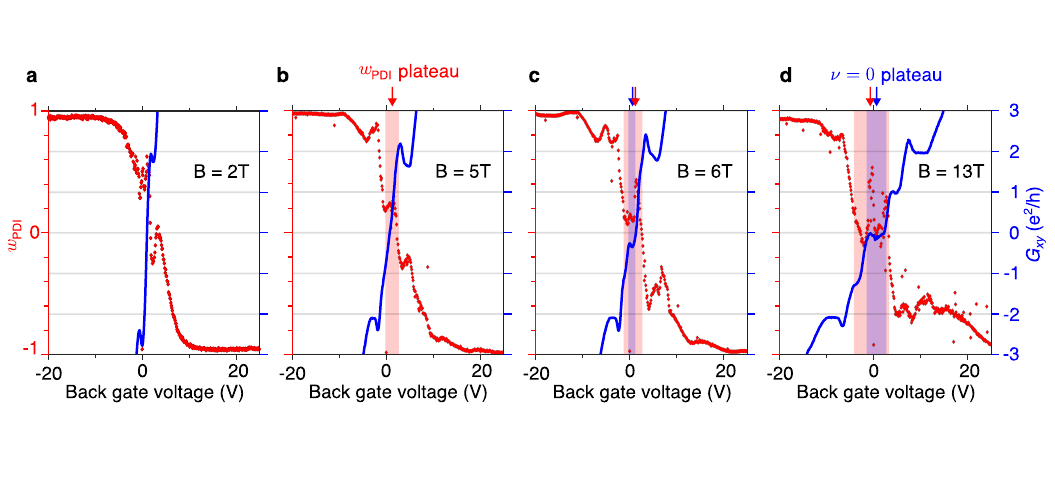}
\caption{
Comparison of $G_{xy}$ (blue) and $w_{\rm PD}$ (red) linecuts at different magnetic fields of \textbf{a:} 2T, \textbf{b:} 5T, \textbf{c:} 6T and \textbf{d:} 13T. 
The areas corresponding to a $\nu = 0$ plateau (condition: $G_{xy} <0.3 e^2/h$) and a $w_{\rm PD}$ plateau (condition: $w_{\rm PD}<0.2$) are highlighted in blue and red, respectively, as a guide for the eye.
}
\label{Fig:Gxy_w_linecuts_comparison}
\end{figure*}

We first observe that $w_{\rm PD}$ is close to the topologically non-trivial values $\pm 1$ for the most part of the ($B$, $V_g$) phase diagram. 
This confirms that, similar to the QH effect in AlGaAs~\cite{Ochkan2024}, the conductance matrix of the QH effect in graphene also corresponds to the HN model and reflects its chirality/non-Hermitian topology, starting at fields as low as 1T. 
Contrary to AlGaAs quantum wells, the carrier density of graphene can easily be tuned using a gate voltage and the charge carrier polarity varied between electron-like and hole-like at a fixed magnetic field. 
Since the topology of the HN model reflects the left-/right-mover chirality, we expect a variation of $w_{\rm PD}$ from $+1$ to $-1$ while varying the chemical potential through the charge neutrality point at constant magnetic field. 
This is indeed observed, as shown in the line cuts presented in Fig.~\ref{Fig:wPDI}(b) and in the red-blue domains of Fig.~\ref{Fig:wPDI}(a) corresponding to the electron- and hole-side, respectively, and as shown by the separation between both at $V_g \simeq 1.9$V, corresponding to charge neutrality.

Interestingly, the best quantization $1-|w_{\rm PD}|$ of the HN topological invariant is observed in regions where the filling factor $\nu = \frac{1}{4}\frac{n h}{e B}$ (with $n$ the carrier density, $h$ and $e$ the Planck constant and electron charge) is very large, corresponding to the low-field/high density region of the fan diagram in Fig.~\ref{Fig:wPDI}(a), rather than on well-quantized QH plateaus like the $\nu=\pm2$ plateaus, as can be seen in Fig.~\ref{Fig:wPDI}(c) (the quantity $1-|w_{\rm PD}|$ measuring the deviation from the perfect quantization). 
Indeed, the deviation of $w_{\rm PD}$ from the $\pm1$ quantized value reaches lower values at constant filling factor when increasing the magnetic field from 1T to 14T.
However, deep in the QH regime, the quantization also increases at higher filling factor for a constant magnetic field, so that better quantization is reached at 3T and $\nu=5$ than at 14T over the whole available $\nu$ range.
This is confirmed by the observation of the non-Hermitian skin effect, which is also most pronounced in the low-field/high $\nu$ regime, and absent in the insulating $\nu=0$ phase (see Supplementary Information and Fig.~\ref{Fig:SI:SPD}).
This result confirms that the well-quantized QH plateaus are not necessary to the global chirality of the system and that a large number of edge channels can be more effective in enforcing chirality. 
The observed quantization does not reach the level reported in AlGaAs \cite{Ochkan2024} ($10^{-2}$ instead of $10^{-4}$). 
We attribute this to the presence of residual conductance channels in the bulk of the graphene, which induce off-diagonal terms in the conductance matrix between non-consecutive contacts. 
Given the lateral sizes of our graphene device ($\simeq 2\mu m$) with respect to that of the AlGaAs one of ref.\cite{Ochkan2024} ($\simeq 100\mu m$), reinforced by the difference in geometry (bar-shape compared to circle), the importance of these finite conductance channels is more visible in graphene.
%We note that the difference of about two orders of magnitude in the quantization of $w_{\rm PD}$ is similar to the ratio of the two devices' widths ($\simeq 2\mu m$ compared to $\simeq 100\mu m$).

The second main feature that can be observed in Fig.~\ref{Fig:wPDI}(a) is the existence of a white area at magnetic fields above 6T, corresponding to $w_{\rm PD}$ close to zero. 
The presence of this feature is made particularly clear in the line cuts of Fig.~\ref{Fig:wPDI}(b), where a widening plateau-like feature around $w_{\rm PD}=0$ (highlighted by the grey region) is observed as the magnetic field is increased.  
This feature occurs close to charge neutrality and around the field at which the insulating $\nu=0$ state, deprived of edge channels, opens. 

To study the $w_{\rm PD}=0$ region in more detail, we show in Fig.~\ref{Fig:Gxy_w_linecuts_comparison} the comparison between $G_{xy}$ and $w_{\rm PD}$ at several magnetic fields.
We identify three different regimes. 
At low field [$B=2$T, Fig.~\ref{Fig:Gxy_w_linecuts_comparison}(a)], no sharp QH plateaus are visible yet, and the HN topological invariant continuously evolves from $-1$ to 1 without a real intermediate plateau-like feature.

At moderate magnetic fields, for $B=5T$, we observe an intermediate regime. 
In this regime, although the $\nu=0$ QH state is absent, the $w_{\rm PD}$ plateau visible at higher field still survives [see Fig.~\ref{Fig:Gxy_w_linecuts_comparison}(b)]. 
Below 5T, this plateau feature in $w_{\rm PD}$ vanishes and only the two satellites are still visible, at $V_g \simeq -1$V and $V_g \simeq +3$V in Fig.~\ref{Fig:Gxy_w_linecuts_comparison}(a) and at $V_g \simeq \pm5$V in Fig.~\ref{Fig:Gxy_w_linecuts_comparison}(b).

We understand the observation of a $w_{\rm PD}$ plateau away from the $\nu=0$ phase as related to a recent theoretical prediction of non-Hermitian topological phases in the conductance matrix of graphene~\cite{Chaturvedi2024}. 
Although in a perfect graphene QH system, $w_{\rm PD}$ would be expected to vary from $-1$ to $+1$ between $\nu=-2$ and $\nu=2$ QH phases, the device imperfections are responsible for the appearance of an additional non-Hermitian phase of the conductance matrix, resulting in a plateau at $w_{\rm PD}=0$ while transiting through charge neutrality. 
In our device, such a transition is only attainable before the first degeneracy lifting of the zeroth Landau level, i.e. at low field in the absence of a $\nu=0$ phase.

At high field ($B=[6-13]$T), corresponding to Fig.~\ref{Fig:Gxy_w_linecuts_comparison}(c-d), the degeneracy of the $N=0$ Landau level is lifted, first partially at $B=6$T, then fully for $B=13$T, and QH plateaus are observed for $\nu=-2,-1,0,1,2$, as can be seen in Fig.~\ref{Fig:Gxy_w_linecuts_comparison}(d). 
Although oscillations of $w_{\rm PD}$ can be seen in the region of the $\nu=1$ and $\nu=2$ QH plateaus (around 5 to 13V), there is no clear evolution of $w_{\rm PD}$ corresponding to the plateau/transition features of the QH effect. 
Around $\nu=0$ however, $w_{\rm PD}$ reaches values close to zero, accompanied by rapid noise. 
%As confirmed by numerical simulations of the winding number of a disordered graphene device (see Supplementary Information), this feature corresponds to the $w_{\rm PD}=0$ plateau recently predicted \cite{Chaturvedi2024} and caused by local disorder and contact imperfections. 
%These fluctuations are probably related to the exact configuration of electron-hole puddles as discussed above, as well as the finite size of the conductance matrix (see SI). 
The same feature is reproduced at $B=6$T [see Fig.~\ref{Fig:Gxy_w_linecuts_comparison}(c)], the lowest field for which a $\nu=0$ plateau is observed, where $w_{\rm PD}$ shows a quasi-plateau close to a zero value.
The observed fluctuations in $w_{\rm PD}$ can be understood as a fingerprint of the exact potential landscape (electron-hole puddles) convoluted by the position and number of the contacts. Increasing the number of contacts reduces the effect of finite size of the conductance matrix, effectively averaging the fluctuations out, as indicated by our numerical simulations (see Supplementary Information and Fig.~\ref{Fig:SI:SPD}). We observe that the width of the $w_{\rm PD}=0$ feature is larger in gate voltage than the $\nu=0$ plateau, as exemplified by the red/blue boxes in Fig.~\ref{Fig:Gxy_w_linecuts_comparison}(c, d) and Supplementary Information Fig.~\ref{Fig:Gxy_w_Fan_comparison}. This might be due to the survival of disorder-induced effects, similar to the lower-field phase. However, the exact correspondence remains to be investigated in details.

The fact that $w_{\rm PD}$ reaches zero in the $\nu=0$ QH phase is due to the lack of chiral edge states. 
At $\nu=0$, no edge channels circulate around the device, so that the only transport occurs through localized QH bulk states. 
Those correspond to balanced electron/hole puddles due to the electrical potential roughness of the graphene layer. 
Since electron and hole puddles are of opposite chirality, the chirality of the global system can be expected to be close to zero. Fluctuations of $w_{\rm PD}$ around zero with gate voltage can be understood as due to local changes of the electron/hole puddles filling (see Supplementary Information).

%\section{Discussion}

%\section{Conclusion}
In conclusion, we investigated the non-Hermitian topology of the HN model in a graphene quantum Hall device. 
We observe that, similarly to what was reported in AlGaAs two-dimensional electron gas, the conductance matrix of a graphene device in the quantum Hall regime has the same form as the the Hamiltonian matrix of the HN model. 
Thanks to the tunability of graphene, the topological character can be tuned between left- and right-chirality of the HN model while passing through charge neutrality at constant magnetic field. 
Interestingly, we find that the best quantization of the topological invariant is observed at high carrier density and low field, rather than on a well quantized plateau such as the $\nu=2$ at highest field. 
This is of particular interest for the optimized operation of devices based on non-Hermitian topology. 
At fields between 5T and 6T, we observe the appearance of a plateau-like feature around $w_{\rm PD}=0$. 
This confirms the recent theoretical prediction of a $w_{\rm PD}=0$ trivial non-Hermitian phase of the conductance matrix outside of the $\nu=0$ quantum Hall phase, caused by imperfections of the device. 
At higher fields, this feature persists in the $\nu=0$ quantum Hall phase of graphene. The relation between the width of the $\nu=0$ phase and that of the $w_{\rm PD}=0$ phase remains to be investigated.% The nature of this spinfull $w_{\rm PD}=0$ phase, potentially distinct from the low-field spin-degenerate one, remains to investigate.

%We further identify another topological phase at $w_{\rm PD}=0$ in the insulating $\nu=0$ quantum Hall phase of graphene. We moreover confirm the recent theoretical prediction of a $w_{\rm PD}=0$ non-Hermitian topological phase of the conductance matrix outside of the $\nu=0$ quantum Hall phase, caused by imperfections of the device.

\paragraph{\bf Acknowledgements} 
This work was supported by the Deutsche Forschungsgemeinschaft (DFG, German Research Foundation) under Germany's Excellence Strategy through the W\"{u}rzburg-Dresden Cluster of Excellence on Complexity and Topology in Quantum Matter -- \emph{ct.qmat} (EXC 2147, 390858490 and 392019). Additionally, E.M.H. acknowledges financial support by SFB1170 ToCoTronics, Project-ID 258499086.
LV was supported by the Leibniz Association through the Leibniz Competition and by the French ANR. 
VK was funded by the European Union.
K.W. and T.T. acknowledge support from the JSPS KAKENHI (Grant Numbers 21H05233 and 23H02052) and World Premier International Research Center Initiative (WPI), MEXT, Japan.

\bibliography{article}% common bib file
\bibliographystyle{apsrev4-2}

\clearpage

\setcounter{figure}{0}
\renewcommand{\thefigure}{S\arabic{figure}}

\section*{Supplementary Information}

\subsection*{Measurement of the resistance and conductance matrices}

\begin{figure*}[t]
\centering
\includegraphics[width=1\textwidth]{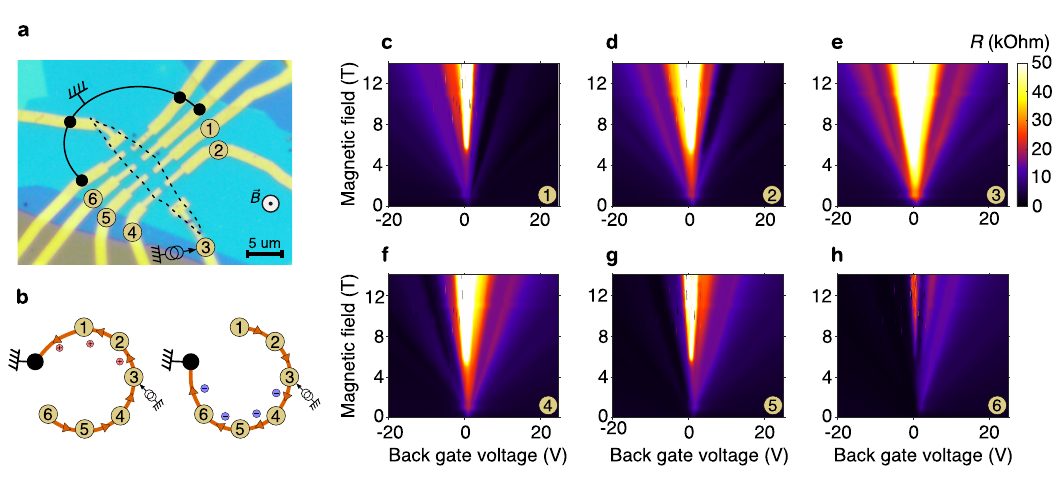}
\caption{
\textbf{Measurement of the resistance matrix. a}: 
schematic of the contact configuration for the measurement of the third column of the resistance matrix.
\textbf{b}: schematic Hatano-Nelson chain of sites corresponding to the contact configuration in the measurement of the third column of the resistance matrix. 
Depending on the chemical potential, the direction of propagation of the edge channels reverses. 
\textbf{c-h}: fan diagram of the resistance $R_{i3} = V_i / I_3$ at contact 1 to 6. %\textcolor{red}{(Cosma: I propose to update the notation such that elements of the conductance matrix $R$ are denoted as $R_{ij}$, with two indexes)}
}
\label{Fig:S_matrix_measurement}
\end{figure*}

In order to obtain the conductance matrix $G$ for a given set of $B$ and $V_g$, we start with the measurement of the resistance matrix $R$ before inverting it to obtain $G$.
Measuring the $j^{\rm th}$ column of the resistance matrix $R$ of the 6-contacts device consists in injecting a current $I_j$ into contact $j$ and measuring the potentials of contacts $i=1$ to $i=6$: $V_i = R_{ij} I_j$ with $R_{ij}$ the elements of the conductance matrix $R$. 
To obtain the resistance and conductance matrices for each point of the ($B$, $V_g$) space, each of these measurements is conducted over the whole fan diagram. 
The fan diagrams corresponding to the measurement of the third column of the resistance matrix ($R_{i3}$, with current injected in contact 3) is displayed in Fig.~\ref{Fig:S_matrix_measurement}.
The fan diagram of contact 3 is symmetric between electron and hole sides, as expected from the source contact. 
The fan diagrams of contacts 1-2 show higher voltage values on the hole side, while contacts 4-5-6 show higher values on the electron side. 
This corresponds to the fact that those contacts are located clockwise (resp. counterclockwise) with respect to the source, so that in the QH regime they equilibrate with either the source potential or with the ground, depending on the chirality of the edge modes. %\textcolor{red}{(Cosma: please check if you agree how I rephrased the sentence. The old sentence is still in the tex file, commented out.)}.
%This correspond to the fact that those contacts are located clockwise (resp. counterclockwise) with respect to the source, so that in the QH regime they equilibrate with the drain potential (ground) in the electron (resp. hole) side, given the clockwise (resp. counterclockwise) direction of propagation of the electron (resp. holes) edge chanels.

\subsection*{Examples of resistance and conductance matrices}

\begin{figure*}[t]
\centering
\includegraphics[width=1\textwidth]{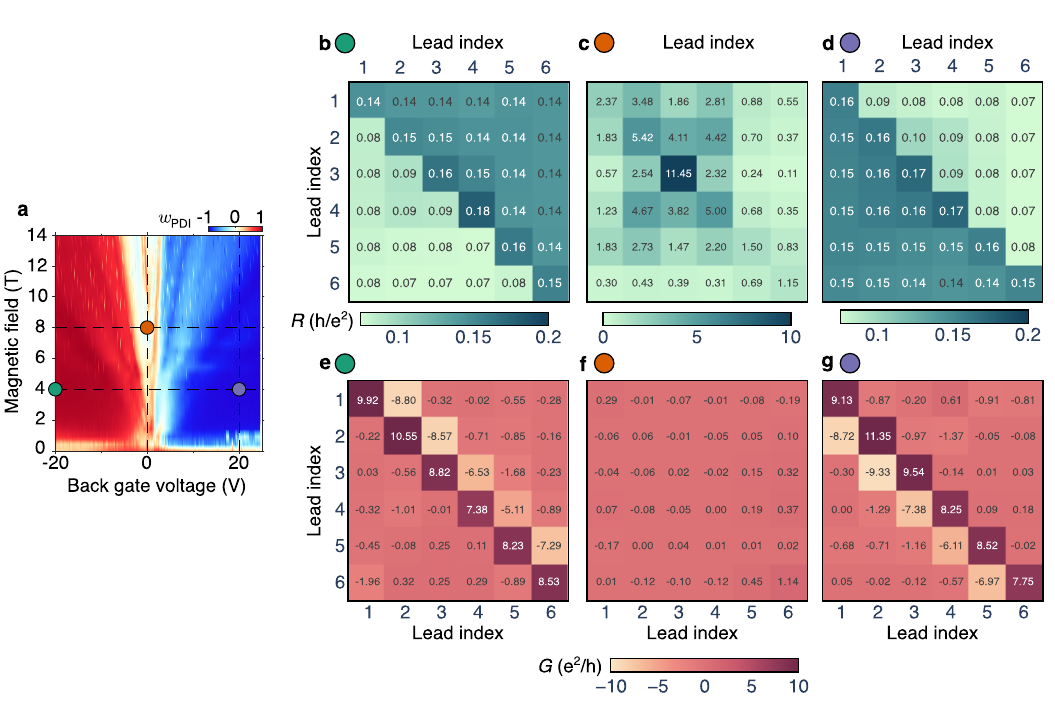}
\caption{
\textbf{Resistance and conductance matrices. a:} Fan diagram of the invariant $w_{\rm PD}$. 
\textbf{b-d}: Resistance and \textbf{e-f} conductance matrices at the points marked in panel \textbf{a}: ($-20$V, 4T) (green), (0V, 8T) (orange) and (20V, 4T)  (purple). 
The rows (resp. the columns) correspond to the index of the current source (resp. voltage probes). 
Each cell of the matrices is color coded from white (min. value) to dark brown (max. value). 
The value of the cell is expressed in units of $G_0 = e^2/h$.
}
\label{Fig:S_matrices}
\end{figure*}

We obtain the conductance matrices by inversion of the real part of the resistance matrices. 
In Fig.~\ref{Fig:S_matrices}, we show some examples of conductance and resistance matrices in the hole-doped, $\nu=0$, and electron-doped regimes in panels Fig.~\ref{Fig:S_matrices}(b, c, d) respectively. 
As expected, the conductance matrices in the heavily-doped regimes are similar to the Hamiltonian of the Hatano-Nelson model in the open-boundary-condition case, with a diagonal of 1 and a sub-diagonal (resp. a super-diagonal) of $-1$ in the hole- (resp. electron-)doped case. On the other hand, the conductance matrix in the $\nu=0$ phase doesn’t show those diagonals, as expected from a diffusive insulating phase dominated by geometrical factors.
%From the measurement of each column of the resistance matrix, we obtain the conductance matrices by inversion of the resistance matrix. 
%In Fig.~\ref{Fig:S_matrices}, we show some examples of conductance and resistance matrices in the hole-doped, $\nu=0$, electron-doped regimes in panels Fig.~\ref{Fig:S_matrices}b,c,d respectively. 
%As expected, the conductance matrices in the heavily-doped regimes are similar to the Hamiltonian of the Hatano-Nelson model in the open-boundary-condition case, with a diagonal of 1 and a sub-diagonal (resp. a super-diagonal) of -1 in the hole- (resp. electron-)doped case. 
%In the $\nu=0$ state, the conductance matrices is mostly homogeneous with very low conductance values, indicating the insulating state and the absence of conducting edge channels.

\subsection*{Invariant calculation} \label{Sec:SM:Invariant_Calculation}

The $w_{\rm PD}$ invariant is a real-space topological invariant that captures the topology of finite-sized non-Hermitian matrices, as previously shown for finite-sized Hamiltonians \cite{Hughes2021} and conductance matrices \cite{Ochkan2024, Chaturvedi2024}. 

A non-zero integer value of this invariant corresponds to a non-trivial topological phase of the matrix, and consequently, the presence of a non-Hermitian skin effect.
The first step in evaluating this for a measured $N \times N$ conductance matrix $G$ is to polar decompose the matrix $\hat{G} = G - \lambda \mathds{1}$, where $\lambda = \text{tr}(G) / N$, tr denotes the trace, and $\mathds{1}$ is the identity matrix. We write $\hat{G}$ as a product of a unitary matrix $Q$ and a positive-definite matrix $P$, $\hat{G} = QP$. 
The invariant is then calculated as:
\begin{equation} \label{eq:wpd_invariant}
    w_{\rm PD}(\hat{G}) = \mathcal{T}(Q^{\dag}[Q, X]),
\end{equation}
where $X={\rm diag}(1, 2, \ldots, N)$ is the position operator encoding the lead index, and $\mathcal{T}$ is the trace per unit volume evaluated over indices away from the ends. 
For the $6 \times 6$ conductance matrices in the main text, this interval consists of indices from 2 to 5.

\subsection*{Numerical simulations}

The fluctuations in the $w_{\rm PD}$ invariant around zero in Fig.~\ref{Fig:Gxy_w_linecuts_comparison} close to the charge neutrality point are a result of finite-size effects due to disorder and to the finite number of leads in the experimental setup.
We show this by simulating the quantum-Hall effect in a nearest-neighbour tight-binding spinless graphene toy model in the presence of a perpendicular magnetic field.
This system has a nearest-neighbour hopping parameter $t = 1$ and chemical potential $\mu$ (in units of $t$), and we construct it in a disk geometry of a fixed radius $r = 240$ (in units of the lattice constant) with $N$ uniformly attached leads (using Kwant \cite{Groth2014}), as described in \cite{Chaturvedi2024}.  

\begin{figure*}[t]
\centering
\includegraphics[width= 0.95\textwidth]{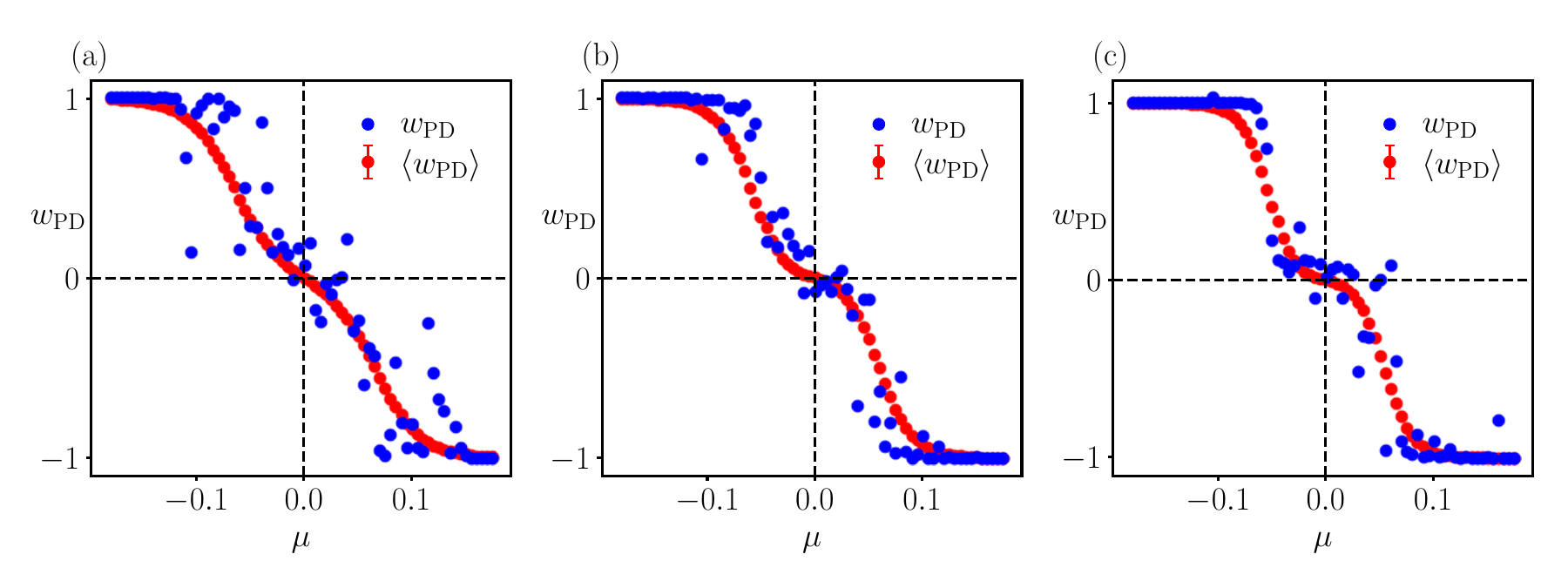}
\caption{
\textbf{Numerical simulation:} $w_{\rm PD}$ (blue) and average $\langle w_{\rm PD} \rangle$ (red) close to the charge neutrality point $\mu = 0$ for different number of leads considered in the transport simulations. Panels (a), (b), (c) correspond to 6, 16, and 26 leads, respectively. Here, the magnetic flux per graphene unit cell is $0.18 \Phi_0$, with $\Phi_0$ the flux quantum, and the disorder strength is $\delta = 1.2$. As the number of leads increases, a clear plateau around 0 emerges for the $w_{\rm PD}$ invariant for the same single disorder realization of the system. However, upon disorder-averaging over multiple disorder configurations of the system, a clear plateau emerges in $\langle w_{\rm PD} \rangle$ even for smaller number of leads in the system.}
\label{Fig:Simulations}
\end{figure*}

In the presence of strong random on-site disorder $\mu \rightarrow \mu + \delta_i$, where $\delta_i$ is chosen from a uniform distribution $(-\delta,\delta)$, the conductance matrix of this system realizes a clear $w_{\rm PD} = 0$ plateau close to the charge neutrality point $\mu = 0$ upon averaging over many disorder configurations of the system (see \cite{Chaturvedi2024}). 
However, for a single disorder configuration of the system, the $w_{\rm PD}$ invariant can have fluctuations, as shown in Fig.~\ref{Fig:Simulations}. 
When $N$ is small, these fluctuations are large because of the interval averaged over in Eq.~\eqref{eq:wpd_invariant} is not sufficiently large. 
However, as the number of leads goes up, this allows for averaging over a larger interval, resulting in the reduction of fluctuations of the $w_{\rm PD}$ invariant. 
On the other hand, the disorder averaged $\langle w_{\rm PD} \rangle$ invariant, which amounts to evaluating the invariant for many disordered realizations of the system, and then finding its mean value, shows an inflection around 0 even for smaller number of leads considered. The results here mimic the case in panel $(a)$ and $(b)$ of Fig.~\ref{Fig:Gxy_w_linecuts_comparison}, before the onset of the $\nu=0$ phase.

\begin{figure*}[t]
\centering
\includegraphics[width=1\textwidth]{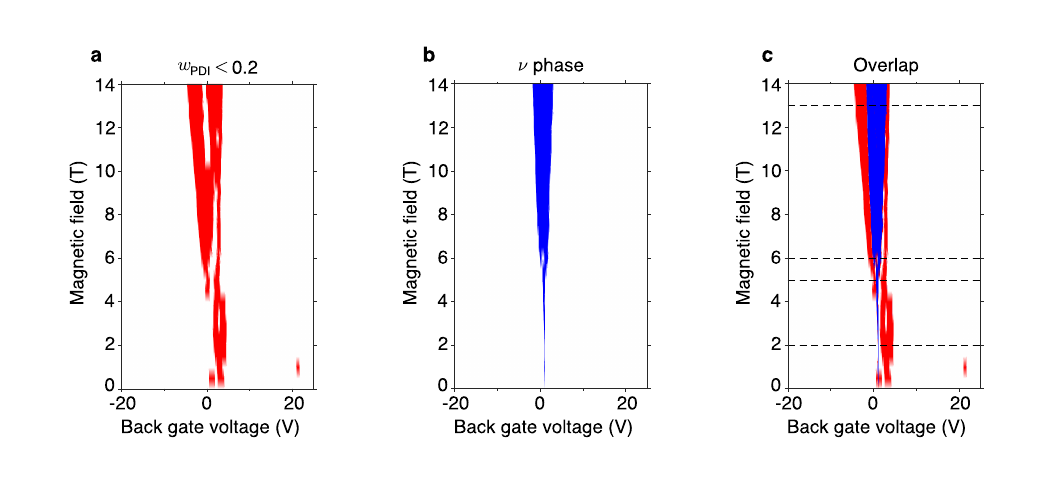}
\caption{
\textbf{Correspondence of the $\nu = 0$ phase and of the $w_{\rm PD} = 0$ area. a}: 
fan diagram of the area where $w_{\rm PD}<0.2$. 
\textbf{b}: fan diagram of the area of the $\nu = 0$ phase (criterion: $G_{xy} <0.3 e^2/h$). 
\textbf{c}: overlap of both areas. 
The dashed lines correspond to the position of the linecuts in Figure.~\ref{Fig:Gxy_w_linecuts_comparison}. 
%The low $w_{\rm PD}$ area reaches significantly beyond the $\nu = 0$ phase, both in field and gate voltage.
}
\label{Fig:Gxy_w_Fan_comparison}
\end{figure*}

\subsection*{Non-Hermitian skin effect}

The non-Hermitian skin effect denotes a phenomenon in which an extensive number of eigenstates are localized at the ends of the system \cite{Yao2018}. 
It can be demonstrated by calculating the sum of probability densities (SPD), which is defined at the $j^{th}$ lattice site as $\sum_i | \bra{r_j} \ket{\Psi_i} |^2$,  where $\ket{\Psi_i}$ are the right eigenvectors of ${\cal H}_{\rm HN}$ and $\ket{r_j}$ denote the lattice site positions. The localization of the eigenstates occurs at the right (left) end of the chain for $|J_{\rm right}|/|J_{\rm left}| > 1(<1)$, and is related to the nontrivial winding of the energy bands \cite{Okuma2020}.

Experimentally, the  sum of probability densities (SPD) is calculated using the components of the $G$ matrix eigenvectors. The SPD at the site $i$ is:
\begin{equation}
  \text{SPD}(i) = \sum_j |V^{j}_{i} |^2,
\end{equation}
where $V^{j}_{i}$ is the $i^\text{th}$ component (related to the site $i$ of the chain) of the $j^\text{th}$ normalized eigenvector $\textbf{V}^{j}$ of the experimentally determined $G$ matrix.

\begin{figure*}[t]
\centering
\includegraphics[width=1\textwidth]{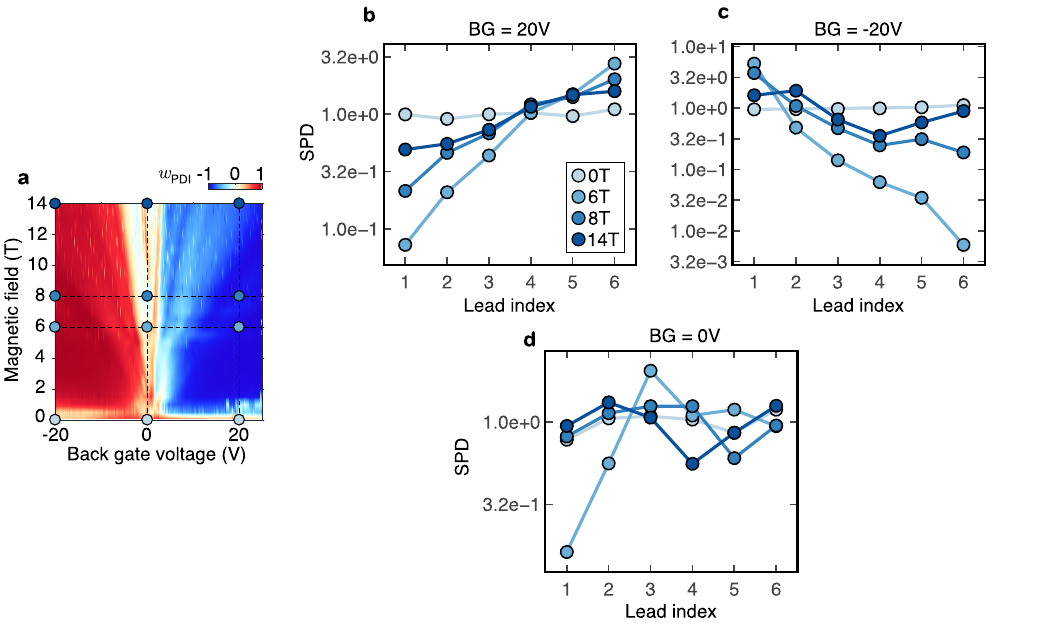}
\caption{
\textbf{Sum of probability densities (SPD) for various values of magnetic field and back gate.}
\textbf{a}: Fan diagram of the invariant $w_{\rm PD}$. 
\textbf{b-c}: SPD in logarithmic scale obtained by diagonalizing the measured $G$ matrix at the data points highlighted in panel \textbf{a}. Highlighted back gate voltages are \textbf{b}: $V=20$~V, 
\textbf{c}: $V=-20$~V, \textbf{d}: $V=0$~V.}
\label{Fig:SI:SPD}
\end{figure*}

Fig.~\ref{Fig:SI:SPD} demonstrates calculated SPDs for a selected number of magnetic fields and back gate voltages. The results coincide with our observations regarding the topological invariant, where the best quantization was obtained at the low-field/high density regions of the fan diagram. Fig.~\ref{Fig:SI:SPD}b,c show the stronger skin effect for magnetic field  value of $6$~T, while as we ramp up the field higher, the non-Hermitian skin effect becomes less pronounced. Additionally, these panels demonstrate how localization occurs on the left or the right ends of the chain depending on the chirality. 
In Fig.~\ref{Fig:SI:SPD}d we can see, as expected for the insulating, $\nu = 0$~QH phase of graphene, no skin effect is observed at any magnetic field value.

\end{document}